\documentclass[conference]{IEEEtran}
\IEEEoverridecommandlockouts

\usepackage{cite}
\usepackage{amsmath,amssymb,amsfonts}
\usepackage{algorithmic}
\usepackage{graphicx}
\usepackage{textcomp}
\usepackage{xcolor}
\usepackage{stfloats}
\newcommand{\real}{\mathbb{R}}

\def\BibTeX{{\rm B\kern-.05em{\sc i\kern-.025em b}\kern-.08em
    T\kern-.1667em\lower.7ex\hbox{E}\kern-.125emX}}

\begin{document}

\title{Data-Driven Post-Event Analysis with Real-World Oscillation Data from Denmark%
\thanks{This work is supported by the Engineering and Physical Science Research Council, UK [grant number EP/Y025946/1].}
}

\author{
\IEEEauthorblockN{Youhong Chen}
\IEEEauthorblockA{%\textit{Electronic and Electrical Engineering} \\
\textit{University of Bath}, Bath, UK \\
yc2930@bath.ac.uk}
\and
\IEEEauthorblockN{Debraj Bhattacharjee \\ Balarko Chaudhuri\\ Mark O'Malley}
\IEEEauthorblockA{%\textit{Electrical and Electronic Engineering} \\
\textit{Imperial College London}, London, UK \\
d.bhattacharjee22@imperial.ac.uk}
\and
\IEEEauthorblockN{Nan Qin\\ Adrian Pilkaer Expethit}
\IEEEauthorblockA{\textit{Energinet}, Fredericia, Denmark \\
naq@energinet.dk;\\ aex@energinet.dk}
}

\maketitle

\begin{abstract}
This paper demonstrates how Extended Dynamic Mode Decomposition (EDMD), grounded in Koopman operator theory, can effectively identify the main contributor(s) to oscillations in power grids. We use PMU data recorded from a real 0.15~Hz oscillation event in Denmark for post-event analysis. To this end, the EDMD algorithm processed only voltage and current phasors from nineteen PMUs at different voltage levels across the Danish grid. In such a blind-test setting with no supplementary system information, EDMD accurately pinpointed the location of the main contributor to the 0.2 Hz oscillation, consistent with the location of the problematic IBR plant later confirmed by Energinet, where the underlying cause was a control system issue. Conventional approaches, such as the dissipating energy flow (DEF) method used in the ISO-NE OSL tool did not clearly identify this plant. This joint validation with Energinet, reinforcing earlier studies using simulated IBR-dominated systems and real PMU data from ISO-NE, highlights the potential of EDMD-based post-event analysis for identifying major oscillation contributors and enabling targeted SSO mitigation.
\end{abstract}


\section{Introduction}
% Sustained low-frequency oscillations have long been a concern in power system operation. Classical inter-area oscillations, arising from electromechanical interactions among synchronous generators in geographically dispersed areas, represent a well-known example. Even under normal operating conditions, routine disturbances may excite such modes. In recent years, however, the power system has undergone a rapid transition towards high penetrations of inverter-based resources (IBRs) for renewable energy integration. This shift has introduced new stability challenges, including the emergence of sub-synchronous oscillations (SSOs) caused by adverse interactions among IBRs through the network. Several recent events worldwide have highlighted the severity of IBR-driven SSOs, compelling system operators to adopt conservative operational strategy and impose limits on IBR output to ensure secure operation.

Poorly damped oscillations induced by inverter-based resources (IBRs) have become a major concern. Identifying the main contributors to such oscillations is essential for effective mitigation. However, this is challenging in IBR-dominated grids because vendor-specific IBR models are opaque, and perturbation-based model estimation is not feasible for post-event analysis.

% The Dissipating Energy Flow (DEF) method has demonstrated strong performance in identifying sources of forced oscillations, particularly those related to synchronous machines \cite{TPWRS21_Slava_DEF}. However, its effectiveness in detecting IBR-driven SSOs remains limited \cite{Fan_Slava_23_DEF_IBR-drivenSSO}. To address this, a modified SSO-CDEF method was proposed in, although its performance remains sensitive to system characteristics. The magnitude of SSO observed at the point of interconnection (PoI) of an IBR plant has also been suggested as an indicator of its participation \cite{dong2023analysisnovember212021}, but the approach lacks rigorous theoretical foundations and is not always reliable. The Grid Impedance Scanning Tool (GIST) developed by NREL \cite{nrel_impedance} can identify IBRs contributing to poorly damped SSO modes, but it relies on dynamic impedance scans using EMT models under multiple operating conditions, which is not suitable for post-event analysis.

The Dissipating Energy Flow (DEF) method performs well for forced oscillations involving synchronous machines \cite{TPWRS21_Slava_DEF} but is less effective for IBR-induced oscillations \cite{Fan_Slava_23_DEF_IBR-drivenSSO}. A modified SSO-Complex Dissipating Energy Flow (SSO-CDEF) method \cite{Fan_Slava_23_DEF_IBR-drivenSSO} has been proposed, though its performance remains system-dependent. Using oscillation amplitude at IBR terminals as an indicator of participation \cite{dong2023analysisnovember212021} lacks theoretical grounding. NREL’s GIST tool \cite{nrel_impedance} relies on electromagnetic transient (EMT)-based impedance scans across multiple operating conditions and is therefore unsuitable for post-event analysis.

Dynamic Mode Decomposition (DMD) is a data-driven, model-free method for extracting spatio-temporal patterns and has recently been applied in power systems for oscillation analysis and control \cite{9780619,9151361,10328697}. However, existing DMD-based approaches have not been validated for IBR-driven oscillations, where dynamics are highly nonlinear and less transparent. To address this, an Extended DMD (EDMD) framework was developed to identify major contributors to oscillations \cite{chen2025datadrivenmethodidentifymajor}. Although \cite{chen2025datadrivenmethodidentifymajor} validated EDMD with simulated data and real data from synchronous machine-based ISO-NE system, it has not been tested on real IBR-driven events. This paper provides the first validation of EDMD using a real oscillation event from the Danish grid. This demonstrates its effectiveness in identifying the main contributors to poorly damped IBR-driven oscillations for post-event analysis.

\section{Methodology}

\subsection{Extended Dynamic Mode Decompression (EDMD)}
This section provides a brief outline of the proposed EDMD-based methodology and further details can be found in \cite{chen2025datadrivenmethodidentifymajor}. The Extended Dynamic Mode Decomposition is the core of this framework. Its essential components and properties are summarised here following the developments in~\cite{williams2015data}. Consider the discrete-time nonlinear system
\begin{align}
x_{k+1} = F(x_k),
\label{eqn: nls}
\end{align}
with $x \in \mathcal{M} \subseteq \real^n$ and $F: \mathcal{M} \rightarrow \mathcal{M}$. The associated Koopman operator $\mathcal{K}$ acts on observables $\psi \in \mathcal{F}$ as
\begin{align}
\mathcal{K}\psi = \psi \circ F,
\label{eqn: Koopman_composition}
\end{align}
yielding the infinite-dimensional linear evolution
\begin{align}
\psi_{k+1} = \mathcal{K}\psi_k.
\label{eqn: Koopman_linear}
\end{align}

These representations describe the same dynamics, and future system states may be expressed through the Koopman mode expansion~\cite{mezic2005spectral}:
\begin{align}
F(x) = \sum_{j=1}^{N_j} \mu_j v_j \varphi_j(x),
\label{eq: KMD}
\end{align}
where $\mu_j$, $\varphi_j$, and $v_j$ denote the Koopman eigenvalues, eigenfunctions, and modes, respectively. Approximating this linear but infinite-dimensional operator from data has become an effective means for analysing nonlinear systems.

EDMD provides such an approximation by projecting the action of $\mathcal{K}$ onto a finite dictionary of observables. Given snapshot pairs $x_i, y_i \in \mathcal{M}$ with $y_i = F(x_i)$, construct
\begin{equation}
X = \begin{bmatrix} x_1 & x_2 & \hdots & x_M \end{bmatrix}, \quad
Y = \begin{bmatrix} y_1 & y_2 & \hdots & y_M \end{bmatrix}.
\label{eqn:dataXY}
\end{equation}
Let $\mathcal{D} = \{ \psi_1, \hdots, \psi_{n_d} \}$ be the dictionary, and define
\begin{align}
\Psi(x) = \big[\psi_1(x); \psi_2(x); \hdots; \psi_{n_d}(x)\big].
\end{align}

The finite-dimensional approximation of the Koopman operator is obtained as
\begin{align}
K \triangleq G^{\dagger}H,
\label{eq:koopmandef}
\end{align}
with
\begin{align}
G = \frac{1}{M}\sum_{j=1}^M \Psi(x_j)^\Psi(x_j),\quad
H = \frac{1}{M}\sum_{j=1}^M \Psi(x_j)^\Psi(y_j).
\label{def_H}
\end{align}
% and $G^{\dagger}$ the pseudo-inverse of $G$.

This completes the concise summary of the standard EDMD procedure used to approximate the Koopman operator in a finite-dimensional subspace. An alternative formulation of the Koopman operator $K$, introduced in~\cite{Klus2016}, defines $ M_K = K^\top$ and is adopted in this work. This representation, also employed in~\cite{Netto2019}, offers a more convenient structure for computing data-driven participation factors and is used here for consistency.

\subsection{Data-driven algorithm}
\subsubsection{Selection of observables}
\label{subsec:Sel_of_Obser}
% The EDMD procedure begins by defining a dictionary of observables that lifts PMU measurements into a higher-dimensional space $\Psi$ to capture nonlinear system behaviour effectively. Following~\cite{9917430}, polynomial and trigonometric functions derived from voltage and current phasors $V\angle\theta_V$ and $I\angle\theta_I$ provide a suitable basis to capture system'snonlinear dynamics. In particular, the active and reactive power,$P = VI\cos(\theta_V-\theta_I), \qquad Q = VI\sin(\theta_V-\theta_I),$ form a quadratic expansion that reflects key nonlinearities. Accordingly, $P$ and $Q$ at the point of interconnection are used as lifted observables in this work. 
The EDMD lifts the voltage and current phasors $V\angle\theta_V$ and $I\angle\theta_I$ from PMUs into a higher-dimensional space $\Psi$ using a dictionary of observables to capture nonlinear system behavior~\cite{9917430}. Polynomial and trigonometric functions of voltage and current phasors, particularly active and reactive power, $P = VI\cos(\theta_V-\theta_I), Q = VI\sin(\theta_V-\theta_I)$
at the point of interconnection are used as lifted observables in this study.

\subsubsection{Filtered mode of interest}
% The dominant poorly damped oscillatory frequency, denoted as $f_s$ is identified from the FFT analysis of the detrended time-series data. ccordingly, the data are filtered using a $4^{\text{th}}$-order Butterworth band-pass filter with cutoff frequencies of$[0.9,\,1.1]f_s$. To mitigate boundary effects, only the central 50–80\% of the filtered signal is retained for analysis. This preprocessing step enhances the accuracy of the EDMD by isolating the dominant oscillatory mode and reducing the required model order in cases where multiple oscillatory components are present.
The dominant poorly damped frequency $f_s$ is identified via FFT. The data are band-pass filtered with a $4^{\text{th}}$-order Butterworth filter at $[0.9,\,1.1]f_s$. To reduce boundary effects, only the central 50–80\% of the filtered signal is used, improving EDMD accuracy by isolating the main mode and lowering the required model order.

\subsubsection{Calculation of reduced-order $\Tilde{M}_K$}
Filtered data are used to compute the finite-dimensional Koopman operator $M_K$ using EDMD. Although an appropriate set of observables spanning a Koopman-invariant subspace would yield an accurate $M_K$, in practice only an approximately invariant set $\Psi(x)$ can be identified. As a result, inappropriate observable selection may introduce spurious eigenvalues, and the dimension of $M_K$ can be large in power system applications. To address this, $M_K$ is projected onto a reduced subspace via truncated SVD of $G = U\Sigma R^{\ast}$. Retaining the $r$ dominant singular values yields $U_r = U(:,1{:}r), \Sigma_r = \Sigma(1{:}r,1{:}r), R_r = R(:,1{:}r),$ and the reduced Koopman operator
\begin{align}
    \Tilde{M}_K = U_r^{\ast} H R_r \Sigma_r^{-1}.
    \label{eqn:hatKmatrix}
\end{align}
This reduced-order operator retains the slow dynamical subspace while suppressing fast transients, aligning naturally with the objective of analysing low-frequency oscillations. The truncation order is determined by detecting elbow points in the singular value curve.

The eigen-decomposition
\begin{align}
    \Tilde{M}_K \Tilde{\Phi} = \Tilde{M}\Tilde{\Phi}
    \label{eqn:eigvals}
\end{align}
provides the reduced eigenvectors ($\Tilde{\Phi}$) and eigenvalues, from which the full-space right and left eigenvectors are reconstructed as
\begin{align}
    \hat{\Phi} = U_r \Tilde{\Phi}, \qquad 
    \hat{\Xi} = \hat{\Phi}^{\dagger}.
    \label{eqn:righteigenvector}
\end{align}

\subsubsection{Data-Driven modal analysis}

The discrete-time eigenvalues $\mu_i$ obtained from~\eqref{eqn:eigvals} are converted to continuous-time values using
\[
\lambda_i = \frac{\ln(\mu_i)}{\Delta t},
\]
from which the damping ratio and oscillation frequency are derived. The participation factors of the observables in mode $i$ are computed from the left and right eigenvectors as~\cite{Netto2019}
\begin{equation}
    p_{si} = \hat{\phi}_{si}\hat{\xi}_{is},
    \label{eq:participation_factor}
\end{equation}
where $p_{si}$ denotes the participation of $s^th$ state variable to mode $i$. The overall participation factors follow from summing over all associated stable variables, and is normalized for comparison across plants.

Therefore, power plants exhibiting large absolute participation factor values are identified as the primary contributors to the poorly damped oscillation.

\section{Case study with EDMD-based method}
\subsection{PMU data from oscillation event in Denmark}
\begin{figure*}[htbp]
  \centering
  \includegraphics[width=\textwidth]{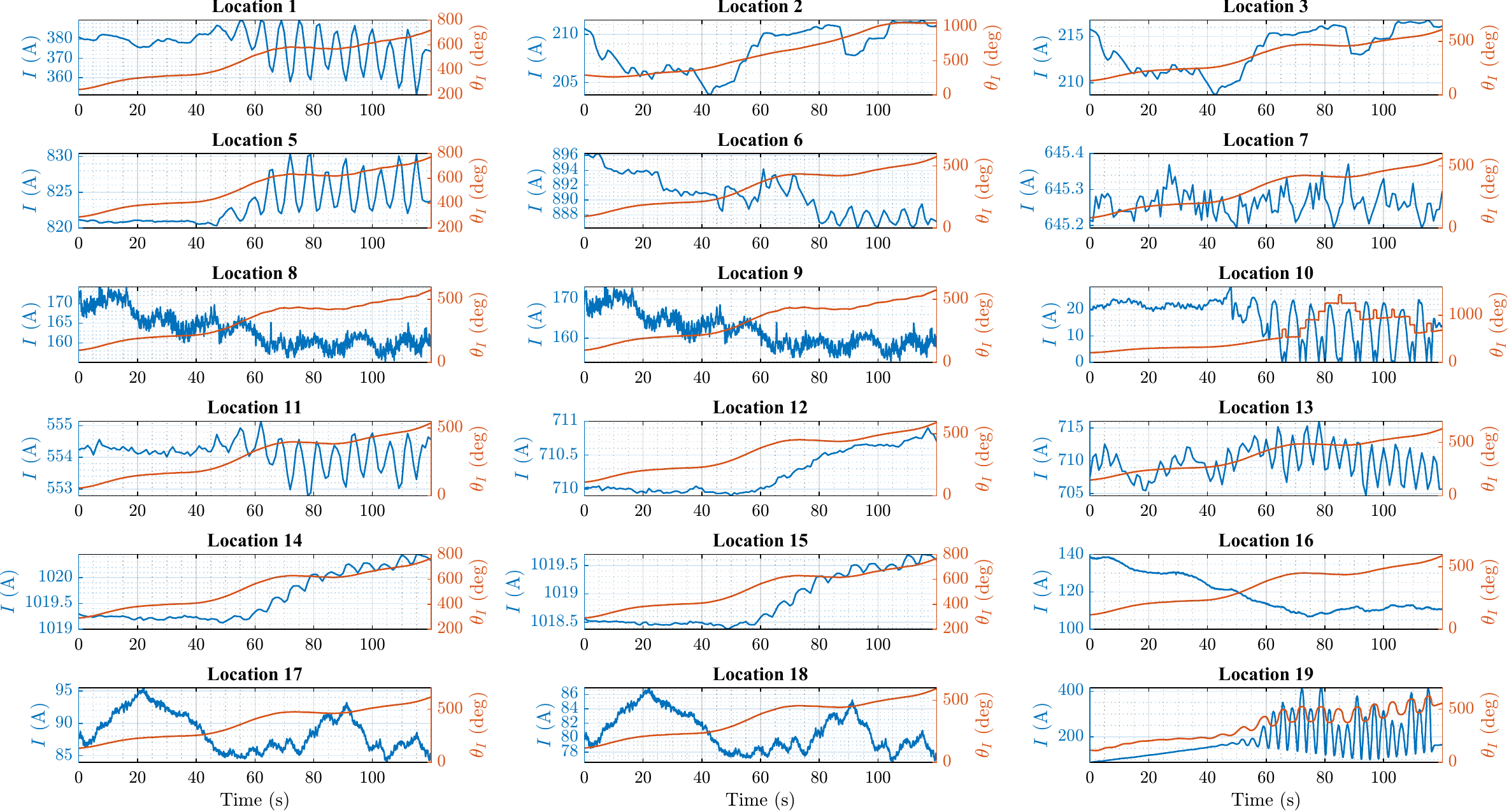}
  \caption{Time‐series phasor current magnitude and angle measurements from PMUs across selected locations in the Danish transmission system.(Location 4 is not included due to bad data quality)}
  \label{fig:current}
\end{figure*}

\begin{figure*}[htbp]
  \centering
  \includegraphics[width=\textwidth]{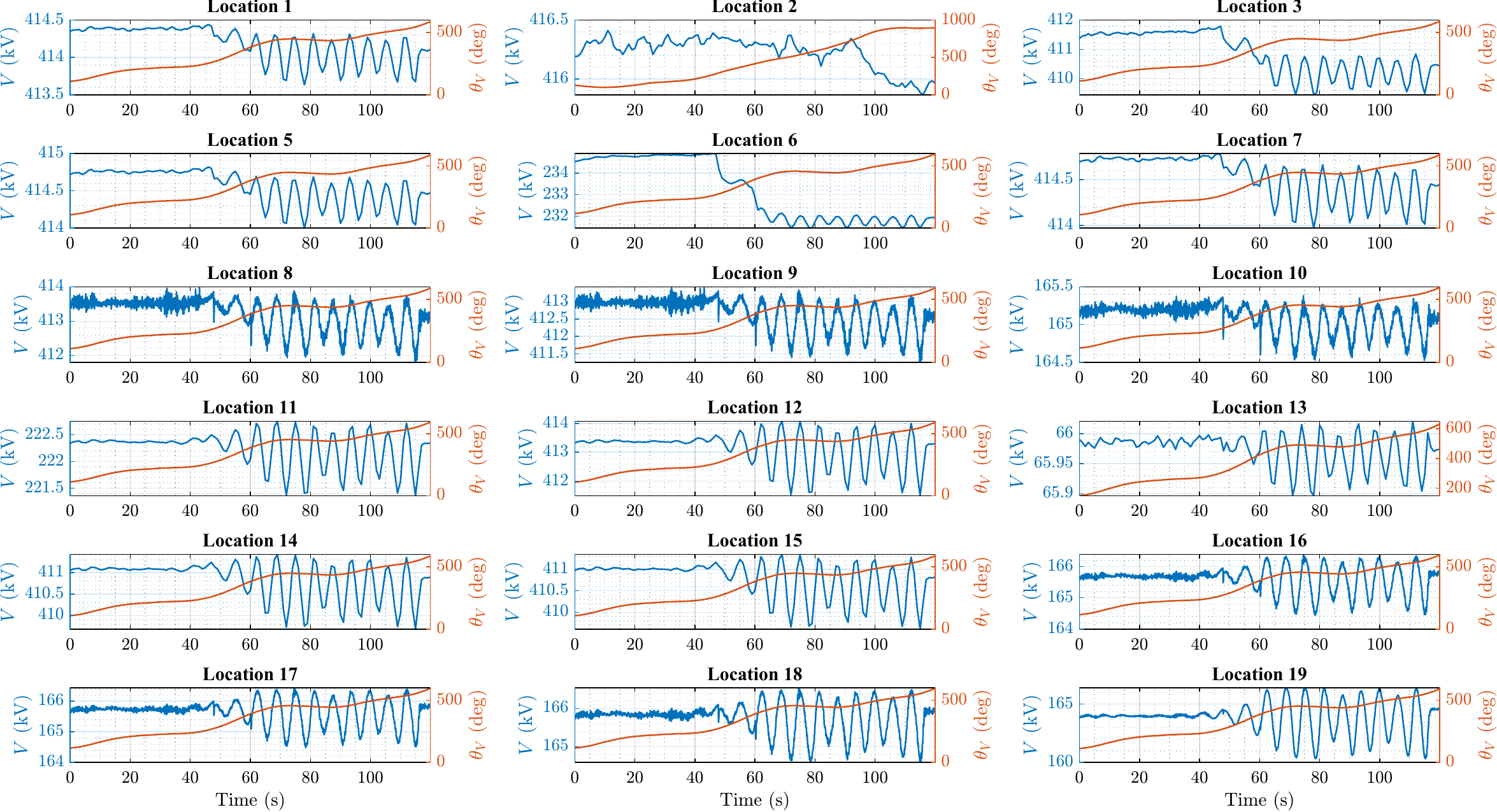}
  \caption{Time‐series phasor voltage magnitude and angle measurements from PMUs across selected locations in the Danish transmission system.}
  \label{fig:voltage}
\end{figure*}

% The Danish transmission network considered in this real system event study comprises 19 locations equipped with Phasor Measurement Units (PMUs). The corresponding PMU data are collected from buses with nominal voltage levels ranging from 60~kV to 400~kV. For each location, both voltage and current phasors are recorded at a reporting rate of 50~Hz. A total of 120 seconds of time-series data are provided, within which pronounced oscillations occur primarily between 60~s and 110~s. Fig.~\ref{fig:current}and \ref{fig:voltage} illustrates the complete 120~s time-series PMU measurements for all locations. For data-driven post-event analysis, the most oscillatory data window (60~s–110~s) is selected.
The oscillation event in Denmark was captured by PMUs at 19 locations with voltage and current phasors recorded at 50 Hz sampling rate. A 120-second dataset is available, within which pronounced oscillations occur between 60–110 seconds. Figures \ref{fig:current} and \ref{fig:voltage} show the full recordings, out of which 60–110 second window is selected for data-driven post-event analysis. Since the dominant oscillation is below 1 Hz, a 3 Hz low-pass filter is applied to suppress high-frequency noise.

% Prior to analysis, the PMU data are pre-processed to remove bad-quality and noisy measurements. Specifically, location 4 is excluded from the dataset due to poor data quality. In addition, as the dominant oscillation frequency of interest is below 1~Hz, a low-pass filter with a cut-off frequency of 3~Hz is applied to suppress high-frequency noise.

\subsection{Observables in EDMD}
\begin{figure*}[htbp]
  \centering
  \includegraphics[width=\textwidth]{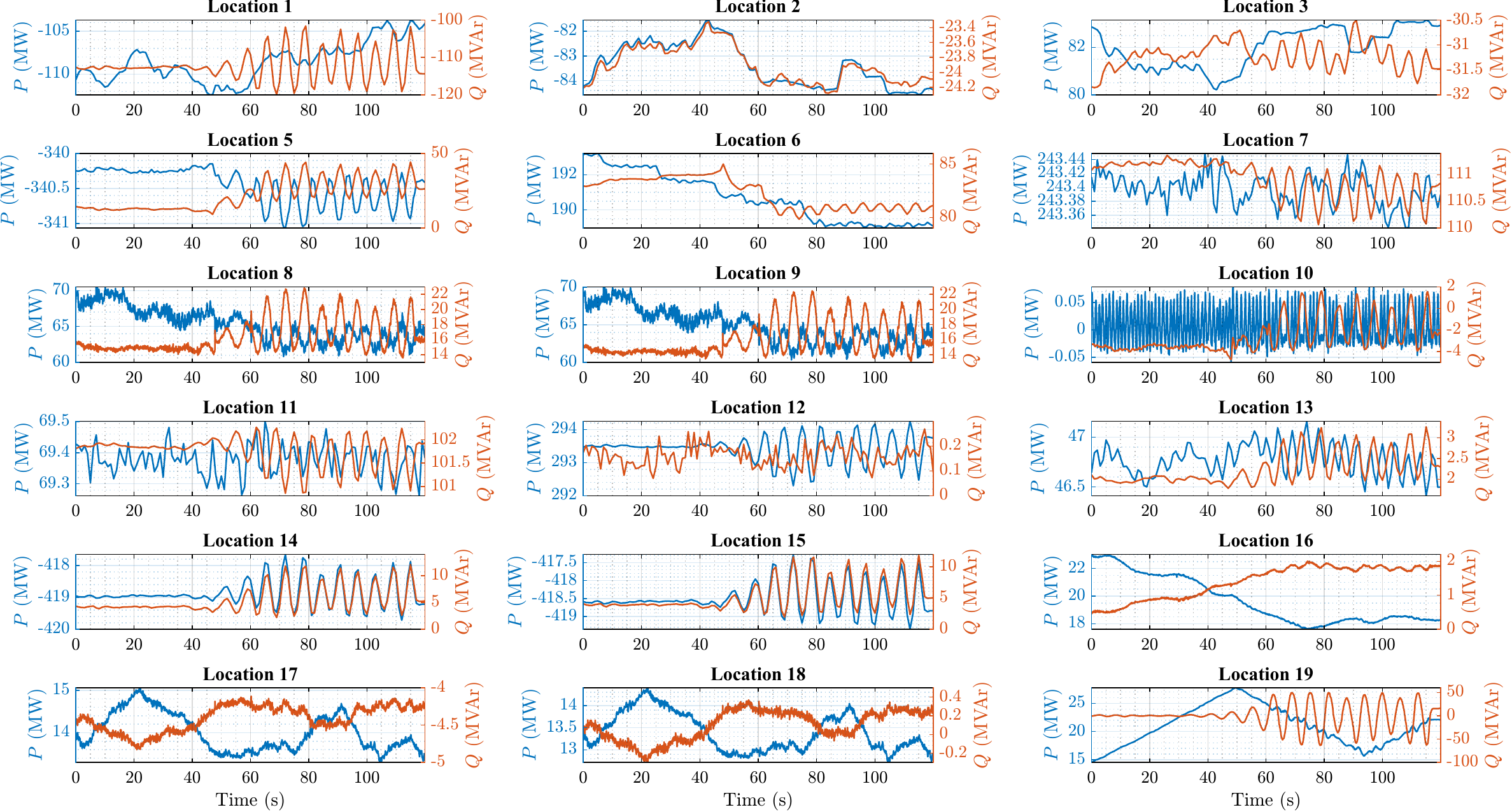}
  \caption{Active and reactive power $P$ and $Q$ calculated from PMU measurements at selected locations in the Danish transmission system.}
  \label{fig:PQ}
\end{figure*}

As discussed in Section~\ref{subsec:Sel_of_Obser}, the active and reactive power components, $P$ and $Q$, which are polynomial functions of the measured voltage and current phasors from PMU data, are more effective in capturing the system nonlinear dynamics. Hence, $P$ and $Q$ are employed in this study. The derived active and reactive power for each location are shown in Fig.~\ref{fig:PQ}.

\subsection{Top oscillation contributor}
% The FFT results reveal a distinct spectral peak at 0.158~Hz, indicating the presence of a dominant oscillatory mode. Based on the singular value analysis, the truncation order for the EDMD analysis is set to 7. The EDMD-based participation factor analysis, presented in Fig.~\ref{fig:dan_pf}, indicates that the generating unit connected at location 19 contributes most significantly to the 0.158~Hz oscillatory mode. According to post-event investigations from Energinet, this mode arose from an inappropriate control setting of a PV plant connected at location 19, triggered by a system fault. The alignment between the EDMD-based findings and Energinet’s post-event investigation confirms the capability of the proposed data-driven EDMD framework to accurately identify the source of oscillatory events in real systems.

\begin{figure}[htbp]
    \centering
    \includegraphics[width=\columnwidth]{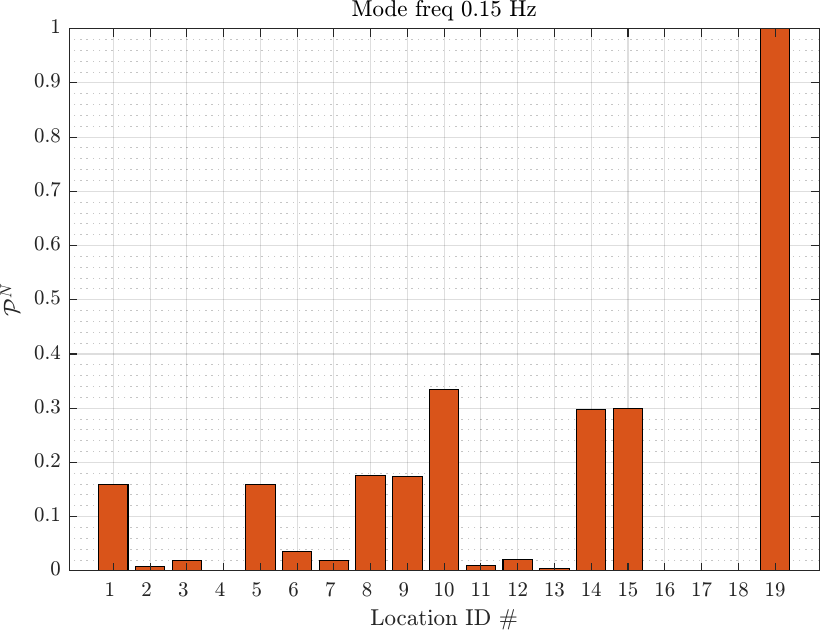} 
    \caption{Data-driven participation factors obtained via EDMD for the Danish transmission system. Location 19 is the top contributor to 0.15 Hz oscillation.}
    \label{fig:dan_pf}
\end{figure}

The FFT reveals a clear spectral peak at 0.158 Hz indicating the dominant oscillatory mode. Singular value analysis yields an EDMD truncation order of 7. The EDMD participation analysis (Fig.~\ref{fig:dan_pf}) identifies location 19 as the main contributor. This is consistent with Energinet’s post-event finding that an IBR plant at location 19 caused the oscillation due to improper control settings. This alignment demonstrates the effectiveness of the proposed EDMD method in pinpointing oscillation contributors from PMU data from a real event.

\section{Case study with other methods}

\subsection{Dissipating Energy Flow (DEF)}
\begin{figure}[htbp]
    \centering
    \includegraphics[width=\columnwidth]{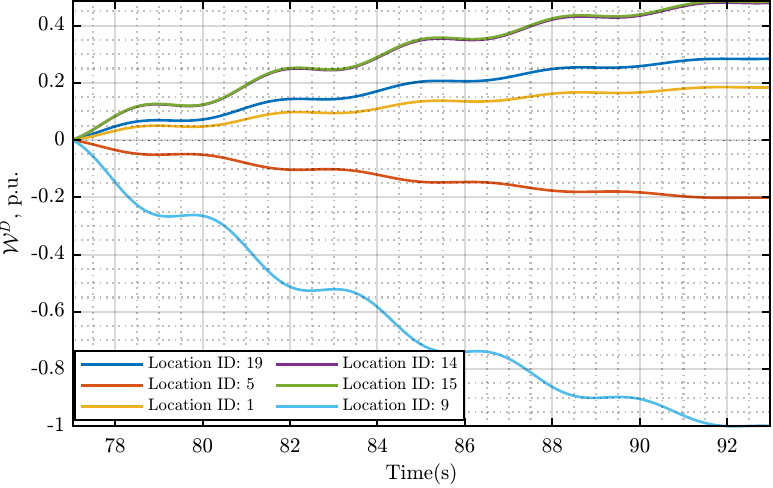} 
    \caption{Top six locations in terms of highest dissipating energy flow in Danish transmission system. Locations 14 and 15 show highest positive energy injection, while location 9 has highest energy absorption.}
    \label{fig:dan_def}
\end{figure}

% For comparison, the Dissipating Energy Flow (DEF) method, which has been widely adopted for oscillation source identification, is also applied to this event. The oscillation source location software (OSLp), developed by ISO New England~\cite{ISO-NE2025software}, based on the DEF algorithm introduced in~\cite{Maslennikov2017}, is employed to analyze the same PMU dataset. Fig.~\ref{fig:dan_def} illustrates the normalized dissipating energy flow over time at each point of interconnection. The DEF results show that locations 14 and 15 exhibit the largest positive dissipating energy (DE) values, followed by locations 19 and 1, while location 5 has a large negative DE value, indicating energy absorption. According to the DEF criterion, sources with large positive DE values are regarded as contributors to the oscillation. Consequently, locations 14 and 15 would be incorrectly identified as the primary oscillation sources, which contradicts the actual post-event investigation.
For comparison, the widely used DEF method is applied using ISO New England’s OSLp software \cite{ISO-NE2025software,Maslennikov2017} on the same PMU dataset from Denmark. Fig.~\ref{fig:dan_def} shows that locations 14 and 15 have the largest positive injecting oscillating energy, while location 9 absorbs energy. Based on the DEF method, locations 14 and 15 would be incorrectly identified as the primary oscillation sources, contradicting the actual post-event findings.

\subsection{Q-V phase-based method}
% Another commonly used approach for oscillation source identification relies on phase relationships between active power and frequency ($P$–$f$) and between reactive power and voltage magnitude ($Q$–$V_m$). If a component’s $P$–$f$ or $Q$–$V_m$ signals are in phase, that component is considered a potential oscillation source. Generally, for synchronous generator (SG)-dominated systems, the $P$–$f$ relationship provides better accuracy, whereas for inverter-based resource (IBR)-dominated systems, the $Q$–$V_m$ relationship is more indicative. Fig.~\ref{fig:pf_qv} presents the time-domain $Q$–$V_m$ relationships for all locations. Locations 7, 11, and 19 exhibit similar in-phase $Q$–$V_m$ patterns. However, this method lacks precision in distinguishing major contributors. Both location 19 (a major contributor) and locations 7 and 11 (minor contributors) are incorrectly identified as top contributors to the poorly damped oscillation. Furthermore, this observation-based technique lacks a rigorous mathematical foundation, limiting its applicability for real power systems.
% Another method used by grid operators relies on phase alignment between $P$–$f$ and $Q$–$V_m$ to identify potential sources of oscillation. For IBR-dominated systems, $Q$–$V_m$ is more indicative. Fig.~\ref{fig:Q_V} shows that locations 7, 11, and 19 exhibit in-phase $Q$–$V_m$ patterns, but the method cannot distinguish major from minor contributors and lacks a rigorous mathematical foundation, limiting its trustworthiness.

\begin{figure*}[htbp]
  \centering
  \includegraphics[width=\textwidth]{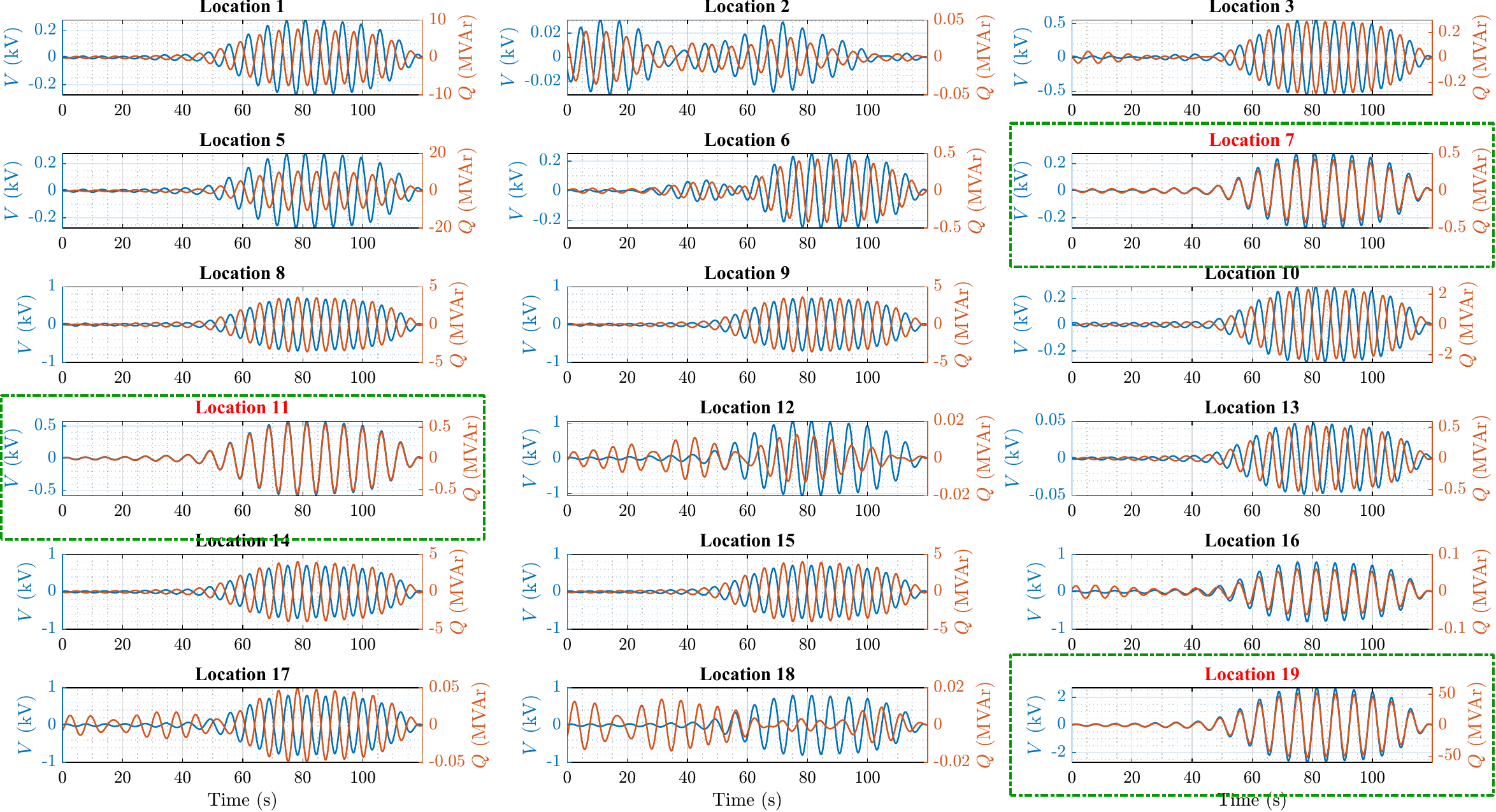}
  \caption{Phase alignment of reactive power $Q$ and voltage magnitude $V$ at the oscillation frequency (band-pass filtered) for selected Danish transmission system locations. Locations 7, 11, 19 show $Q$-$V$ in-phase (marked with a green box).}
  \label{fig:Q_V}
\end{figure*}

Energinet identifies the main oscillation contributors primarily by determining the phase alignment between $Q$–$V$ at the oscillation frequency. Fig.~\ref{fig:Q_V} shows that locations 7, 11, and 19 exhibit in-phase $Q$–$V$ patterns. When combined with practical information (e.g., new plants undergoing testing), this approach suggests location 19 as the major contributor. While this method is effective for detecting forced oscillations, as in the case presented, it is less reliable for natural oscillations because the results are highly sensitive to the selected window.

\section{Conclusion}
This paper demonstrates that Extended Dynamic Mode Decomposition (EDMD) can effectively identify the dominant contributor to oscillations using limited PMU data, without any network information. Its successful application to a real 0.2 Hz oscillation event in Denmark, carried out as a blind test, confirms its accuracy in real-world post event analysis. Unlike conventional dissipating energy flow (DEF) or phase-based methods, EDMD correctly reveals the top contributor to the oscillation even in IBR-dominated grids. These results highlight EDMD’s potential as a powerful data-driven tool for post-event event analysis to ensure the effective mitigation.

\bibliographystyle{IEEEtran}
% argument is your BibTeX string definitions and bibliography database(s)
\bibliography{EDMD_paper_refs}
\end{document}